\documentclass[prd,preprint,superscriptaddress,showpacs,byrevtex]{revtex4}
\usepackage{epsfig}
\newcommand{\be}{\begin{equation}}
\newcommand{\ee}{\end{equation}}
\newcommand{\bea}{\begin{eqnarray}}
\newcommand{\eea}{\end{eqnarray}}

\begin{document}

\begin{flushright}
YITP-SB-06-40
\end{flushright}

\title{\noindent Shift Theorem Involving the Exponential of a Sum of Non-Commuting Operators
in Path Integrals }

\author{Fred Cooper} \email{fcooper@nsf.gov}
\affiliation{Physics Division, National Science Foundation, Arlington VA 22230}

\author{Gouranga C. Nayak} \email{nayak@max2.physics.sunysb.edu}
\affiliation{ C. N. Yang Institute for Theoretical Physics, Stony Brook University, SUNY, Stony Brook,
NY 11794-3840, USA }

\date{\today}

\begin{abstract}

We consider expressions of the form of  an exponential of  the sum
of two non-commuting operators of a single variable inside a path integration.
We show that it is possible to shift one of the non-commuting operators
from the exponential to other functions which are pre-factors and post-factors when the domain of integration of the
argument of that function is from $-\infty$ to $+\infty$. This shift
theorem is useful to perform certain integrals and path integrals involving
the exponential of sum of two non-commuting operators.

\end{abstract}
\maketitle

\newpage

\section{Introduction}

In background field methods in quantum field theory
one often encounters the exponential of the  sum of non-commuting
operators inside the path integration.
A simple example of this type occurs when looking at pair production of  charged scalars \cite{schwinger} \cite{gouranga}
 in the presence of a time dependent background  electric field $E(t)$ in the longitudinal ( $z$)  direction.
In the Axial gauge  $A_z=0$  so that
\bea
A_0 = -E(t)z.
\label{7}
\eea
The action can be written in the form  \cite{cg}
\bea
&& S^{(1)}=-i \int_0^\infty \frac{ds}{s}
\int_{-\infty}^{+\infty} dt <t| \int_{-\infty}^{+\infty} dx
<x| \int_{-\infty}^{+\infty} dy <y| \int_{-\infty}^{+\infty} dz <z| \nonumber \\
&& [ e^{-is[(\hat{p}_0+eE(t) z)^2-\hat{p}_z^2-\hat{p}_T^2-m^2-i\epsilon]} -
e^{-is(\hat{p}^2-m^2-i\epsilon)}] |z> |y> |x> |t>.
\label{10}
\eea
Inserting complete set of $|p_T>$ states $\int d^2 p_T |p_T><p_T|~=~1$
we find (we use the normalization $<q|p>=\frac{1}{\sqrt{2\pi}} ~e^{iqp}$)
\bea
&& S^{(1)}=\frac{-i }{(2\pi)^2}
\int_0^\infty \frac{ds}{s} \int d^2x_T \int d^2p_T
e^{is(p_T^2+m^2+i\epsilon)}[\int_{-\infty}^{+\infty} dt <t|
\int_{-\infty}^{+\infty} dz <z| \nonumber \\
&& e^{-is[(-i\frac{d}{dt}+eE(t) z)^2-\hat{p}_z^2]} |z> |t>- \int dt \int dz
\frac{1}{4\pi s}].
\label{11}
\eea
Inserting $|p_z>$ and $|p_0>$ complete set of states we get
\bea
&& S^{(1)}=\frac{-i }{(2\pi)^2}
\int_0^\infty \frac{ds}{s} \int d^2x_T \int d^2p_T \int dp_0 \int dp'_0 \int dp_z \int dp'_z
e^{is(p_T^2+m^2+i\epsilon)}[\int_{-\infty}^{+\infty} dt ~e^{itp_0}
\int_{-\infty}^{+\infty} dz ~e^{izp_z} \nonumber \\
&& e^{-is[(-i\frac{d}{dt}+eE(t) z)^2-\hat{p}_z^2]} ~e^{-izp'_z} ~e^{-itp'_0}- \int dt \int dz
\frac{1}{4\pi s}].
\label{11}
\eea

In the coordinate
representation the operators ${\hat p}_0~=~\frac{1}{i} \frac{\partial}{\partial t }$
and $E(t)$ do not commute with each other.  In order to evaluate this type of Path  Integral
 it is quite useful to be able to shift the  derivative operator from the exponential
to pre-factor and post-factor  functions that occur when we insert complete sets of states in order to evaluate the Path Integral.

In particular we would like to show that the following theorem is true:
\bea
&& \int_{-\infty}^{+\infty} dy \int_{-\infty}^{+\infty} dx ~
<y|~<x|~
e^{-[(a(y)x+h\frac{d}{dy})^2~+b\frac{d^2}{dx^2}~+c(y)]}~
|x>~|y>~= \nonumber \\
 && \int_{-\infty}^{+\infty} dy \int_{-\infty}^{+\infty} dx ~<y|~<x-\frac{h}{a(y)}\frac{d}{dy}|~
e^{-[a^2(y)x^2~+~b\frac{d^2}{dx^2}~+c(y)]}~
|x-\frac{h}{a(y)}\frac{d}{dy}>~|y> \nonumber \\
\label{ef3xccx}
\eea
which after inserting complete set of states can be written as
\bea
&& \int_{-\infty}^{+\infty} dy \int_{-\infty}^{+\infty} dx ~
<y|~<x|~
e^{-[(a(y)x+h\frac{d}{dy})^2~+~b\frac{d^2}{dx^2}~+c(y)]}~
|x>~|y>= \frac{1}{(2\pi)^2}\int dp_y \int dp'_y \int dp_x \nonumber \\
&&~\int dp'_x \int_{-\infty}^{+\infty} dy \int_{-\infty}^{+\infty} dx
e^{iyp_y}~e^{ip_x(x-\frac{h}{a(y)}\frac{d}{dy})} <p_y|<p_x| e^{-a^2(y)x^2+b\frac{d^2}{dx^2}
+c(y)}|p'_x>|p'_y> \nonumber \\
&& e^{-ip'_x(x-\frac{h}{a(y)}\frac{d}{dy})}e^{-iyp'_y}~~~~~
\label{ef3x}
\eea
where $x$ integration from $-\infty$ to $+\infty$ must be performed for eqs. (\ref{ef3xccx}) and
(\ref{ef3x}) to be true. Here $h$ (which is equal to $i$ in most of the physical examples, see eq. (\ref{11}))
and $b$ are constants and $a, c$ are functions of single variable,
such that the integration over $x$ is well defined. In what follows we
will assume that $a(y)$ and $c(y)$ are sufficiently differentiable, integrable etc.
so that all the formal manipulations are valid. We have used the normalization
\bea
<x|p_x>=\frac{1}{\sqrt{2\pi}} e^{i x p_x}.
\eea
It can be noted that eq. (\ref{ef3x}) can not be derived by replacing
\bea
x \rightarrow x -\frac{h}{a(y)} \frac{d}{dy}
\label{pf}
\eea
directly in eq. (\ref{ef3x}).
This is because $a(y)$ and $x$ commute with each other in the exponential
whereas $a(y)$ and $\frac{d}{dy}$ do not commute with each other under this
replacement. Hence we will use a similarity transformation technique
to derive the above theorem which avoids this problem.
The shift theorem leads to the special case
\bea
&&W= \int_{-\infty}^{+\infty} dy \int_{-\infty}^{+\infty} dx <y|~<x|e^{-[(a(y)x+h\frac{d}{dy})^2]}~|x> ~|y> \nonumber \\
&&~=\frac{1}{(2\pi)^2} \int_{-\infty}^{+\infty}dy~\int_{-\infty}^{+\infty} dx ~\int dp_y \int dp'_y \int dp_x \int dp'_x~e^{ixp_x}~e^{iyp_y}<p_y|e^{-ip_x\frac{h}{a(y)}\frac{d}{dy}} \nonumber \\ && <p_x|e^{-a^2(y)x^2}|p'_x>e^{ip'_x\frac{h}{a(y)}\frac{d}{dy}}|p'_y>~e^{-ixp'_x}~e^{-iyp'_y}.
\label{lr}
\eea
Let us evaluate the left and right hand side of the above equation separately.
For the left hand side we find
\bea
&&W= \int_{-\infty}^{+\infty} dy \int_{-\infty}^{+\infty} dx <y|~<x|e^{-[(a(y)x+h\frac{d}{dy})^2]}~|x> ~|y> \nonumber \\
&&= \int_{-\infty}^{+\infty} dy \int_{-\infty}^{+\infty} dx \int dp_y \int dp'_y <y|p_y>~<x|x><p_y|e^{-[(a(y)x+h\frac{d}{dy})^2]}|p'_y> <p'_y|y>. \nonumber \\
\eea
Since $<p_y|e^{-[(a(y)x+h\frac{d}{dy})^2]}|p'_y>$ is independent of $\frac{d}{dy}$ we can take $<p'_y|y>$ to the left. We find
\bea
&&W= \int_{-\infty}^{+\infty} dy \int_{-\infty}^{+\infty} dx \int dp_y \int dp'_y \int dp_x <p'_y|y><y|p_y><x|p_x><p_x|x> \nonumber \\
&& <p_y|e^{-[(a(y)x+h\frac{d}{dy})^2]}|p'_y> \nonumber \\
&&= \frac{1}{2\pi} \int dp_x \int dp_y <p_y|\int_{-\infty}^{+\infty} dx ~e^{-(a(y)x+h\frac{d}{dy})^2}|p_y>
\label{left}
\eea
Although eq. (\ref{left}) is formally infinite, the $x$-integral
inside $W$
\bea
I(y)= \int_{-\infty}^{+\infty} dx ~e^{-(a(y)x+h\frac{d}{dy})^2~}
\label{xint}
\eea
is finite.

Now evaluating the right hand side of eq. (\ref{lr}) we find
\bea
&&~W
=\frac{1}{(2\pi)^2} \int_{-\infty}^{+\infty}dy~\int_{-\infty}^{+\infty} dx ~\int dp_y \int dp'''_y \int dp_x \int dp'_x~e^{ixp_x}~e^{iyp_y}<p_y|e^{-ip_x\frac{h}{a(y)}\frac{d}{dy}} \nonumber \\
&& <p_x|e^{-a^2(y)x^2}|p'_x>e^{ip'_x\frac{h}{a(y)}\frac{d}{dy}}|p'''_y>~e^{-ixp'_x}~e^{-iyp'''_y} \nonumber \\
&& =\frac{1}{(2\pi)^2} \int_{-\infty}^{+\infty}dy~\int_{-\infty}^{+\infty} dx ~\int dp_y \int dp'_y \int dp''_y \int dp'''_y
\int dp_x \int dp'_x~e^{iyp_y}~e^{ixp_x}<p_y|e^{-ip_x\frac{h}{a(y)}\frac{d}{dy}}|p'_y> \nonumber \\
&& <p_x|<p'_y|e^{-a^2(y)x^2}|p''_y>|p'_x>~<p''_y|e^{ip'_x\frac{h}{a(y)}\frac{d}{dy}}|p'''_y>~
e^{-ixp'_x}~e^{-iyp'''_y}.
\label{eff3}
\eea
It can be shown that
\bea
<p_y|f(y)|p'_y>=\int dy' <p_y|f(y)|y'><y'|p'_y>= \frac{1}{(2\pi)}\int dy' e^{iy(p_y-p'_y)}f(y')
\eea
is independent of $y$ and
\bea
<p_y|f(y)\frac{d}{dy}|p'_y>=-i\int dy' <p_y|f(y)|y'><y'|p'_y>p'_y=-ip'_y\frac{1}{(2\pi)} \int dy' e^{iy(p_y-p'_y)}f(y') \nonumber \\
\eea
is independent of $y$ and $\frac{d}{dy}$.
Hence we can easily integrate over $y$ in eq. (\ref{eff3}). Also since $<p''_y|e^{ip'_x\frac{h}{a(y)}\frac{d}{dy}}|p'''_y>$ is independent of $\frac{d}{dy}$ we can
bring it to the left. We find from eq. (\ref{eff3})
\bea
&&~W
=\frac{1}{(2\pi)}\int_{-\infty}^{+\infty} dx ~\int dp_y \int dp'_y \int dp''_y
\int dp_x \int dp'_x~e^{ix(p_x-p'_x)} \nonumber \\
&& <p''_y|e^{ip'_x\frac{h}{a(y)}\frac{d}{dy}}|p_y><p_y|e^{-ip_x\frac{h}{a(y)}\frac{d}{dy}}|p'_y> <p_x|<p'_y|e^{-a^2(y)x^2}|p''_y>|p'_x> \nonumber \\
&& =\frac{1}{(2\pi)}\int_{-\infty}^{+\infty} dx ~\int dp_y \int dp'_y \int dp''_y
\int dp_x \int dp'_x~e^{ix(p_x-p'_x)} \int_{-\infty}^{+\infty} dx' \nonumber \\
&& <p''_y|e^{ip'_x\frac{h}{a(y)}\frac{d}{dy}}|p_y><p_y|e^{-ip_x\frac{h}{a(y)}\frac{d}{dy}}|p'_y> <p'_y|
<p_x|e^{-a^2(y)x^2}|x'><x'|p'_x>|p''_y> \nonumber \\
&& =\frac{1}{(2\pi)^2}\int_{-\infty}^{+\infty} dx ~\int dp_y \int dp'_y \int dp''_y
\int dp_x \int dp'_x~e^{ix(p_x-p'_x)} \int_{-\infty}^{+\infty} dx' e^{ix'(p_x-p'_x)}\nonumber \\
&& <p''_y|e^{ip'_x\frac{h}{a(y)}\frac{d}{dy}}|p_y><p_y|e^{-ip_x\frac{h}{a(y)}\frac{d}{dy}}|p'_y> <p'_y| e^{-a^2(y){x'}^2}
|p''_y>.
\label{5}
\eea
Since the $x$ dependence is only in $e^{ix(p_x-p'_x)}$ we can now easily integrate over $x$ to find
\bea
&& W =\frac{1}{(2\pi)} ~\int dp_y \int dp'_y \int dp''_y
\int dp_x \int dp'_x~ \int_{-\infty}^{+\infty} dx' \nonumber \\
&& <p''_y|e^{ip_x\frac{h}{a(y)}\frac{d}{dy}}|p_y><p_y|e^{-ip_x\frac{h}{a(y)}\frac{d}{dy}}|p'_y> <p'_y| e^{-a^2(y){x'}^2}
|p''_y> \nonumber \\
&& = \frac{1}{(2\pi)} ~\int dp'_y \int dp''_y \int dp_x \int dp'_x~ \int_{-\infty}^{+\infty} dx' <p''_y|p'_y> <p'_y| e^{-a^2(y){x'}^2}
|p''_y> \nonumber \\
&&~=\frac{1}{2\pi} \int_{-\infty}  dx' \int dp'_y \int dp_x <p'_y|e^{-a^2(y){x'}^2}|p'_y> \nonumber \\
&& =\frac{1}{2\pi} \int dp_y \int dp_x <p_y| \int_{-\infty}^{+\infty}dx ~e^{-a^2(y)x^2}|p_y> \nonumber \\
\label{eff3f}
\eea

Although eq. (\ref{eff3f}) is formally infinite, the $x$-integral
inside $W$
\bea
I(y)= \int_{-\infty}^{+\infty} dx ~e^{-a^2(y)x^2}
\label{xint1}
\eea
is finite.

Hence from eqs. (\ref{left}) and (\ref{eff3f}) we find
\bea
I(y)= \int_{-\infty}^{+\infty} dx ~e^{-(a(y)x+h\frac{d}{dy})^2~}=\int_{-\infty}^{+\infty} dx ~e^{-a^2(y)x^2}.
\label{special}
\eea

This above theorem  eq. (\ref{ef3x})  can be generalized to involve matrices as follows
\bea
&& I_{ij}(y)~=
~[\int_{-\infty}^{+\infty} dy \int_{-\infty}^{+\infty} dx <y|~<x|e^{-[(A(y)x+h\frac{d}{dy})^2~+B\frac{d^2}{dx^2}~+
C(y)]}~|x> ~|y>]_{ij} \nonumber \\
&&~= \frac{1}{(2\pi)^2}[\int_{-\infty}^{+\infty} dy~\int_{-\infty}^{+\infty} dx ~\int dp_y \int dp'_y \int dp_x
\int dp'_x~ e^{iyp_y}~e^{ixp_x} <p_y|e^{-ip_x\frac{h}{A(y)}\frac{d}{dy}}  \nonumber \\
&& <p_x|e^{-[A^2(y)x^2~+~B\frac{d^2}{dx^2}~+C(y)]}|p'_x>e^{ip'_x\frac{h}{A(y)}\frac{d}{dy}}|p'_y>~
~e^{-ixp'_x}e^{-iyp'_y}]_{ij}.
\label{ef4x}
\eea
Here $h$ and $B$ are constants and $A^{ij}(y)$, $C^{ij}(y)$,
are ($i,~ j$) dimension matrices which do not commute with $\frac{d}{dy}$;
chosen that the integration over $x$ is well defined. In what follows we
will assume that $A^{ij}(y)$ and $C^{ij}(y)$ are sufficiently differentiable, integrable etc.
so that all the formal manipulations are valid.

This shift by derivative technique will be very useful when one studies particle production
from arbitrary background fields via Schwinger-like mechanisms in QED and QCD \cite{schwinger,gouranga}.
Quark and gluon production from arbitrary classical chromofields is expected to be an important ingredient
in the production and equilibration of the quark-gluon plasma found at the RHIC and LHC \cite{cooper,gouranga1}.

This paper is organized as follows. In section II we provide general derivations of eqs. (\ref{ef3x}), and (\ref{ef4x})
by using similarity transformation techniques. We verify eq. (\ref{special}) by directly performing the integration, where
we consider the integrals as a function of the variable $h$ and assume that the integrals have a unique Taylor Series
in $h$. We present our conclusions in section IV.

\section{Similarity Transformation Approach for  Deriving  the ``Shift Theorem"}

In this section we provide a general derivation
of eqs. (\ref{ef3x}) and (\ref{ef4x})
by using similarity transformations. Before giving such a
derivation we consider here a simple case ($a(y)=a$=constant)
to show how the derivative operator acts as a c-number
when $x$ integration is from $-\infty$ to $+\infty$.
We find by using Fourier transformation technique
\bea
&& \int_{-\infty}^{+\infty} dx e^{-(ax+h\frac{d}{dy})^2} f(y)
=\int_{-\infty}^{+\infty} dx e^{-(ax+h\frac{d}{dy})^2} \int dp f(p) e^{iyp}
=\int dp \int_{-\infty}^{+\infty} dx e^{-(ax+ihp)^2} f(p) e^{iyp} \nonumber \\
&& =\int dp \int_{-\infty}^{+\infty} dx e^{-a^2x^2} f(p) e^{iyp}
= \int_{-\infty}^{+\infty} dx e^{-a^2x^2} f(y)
\eea
In the above we have assumed that $f(y)$ is well enough behaved so that it is legal to change the order in which the integrations are taken.  In what follows we will assume that $f(y)$ is sufficiently differentiable, integrable etc. so that all the formal manipulations are valid.

However, this Fourier transformation
technique does not work if $a(y)$ is not a constant. This is because
$a(y)$ and
$\frac{d}{dy}$ do not commute with each other in the exponential.
For this purpose we use a similarity transformation technique to derive the shift theorem.

\subsection{Shift Theorem Involving Non-Commuting Operators in the Exponential}

Consider the following similarity transformations acting  on $x$:
\bea
x \pm \frac{h}{a(y)}\frac{d}{dy}~=~
e^{ \pm \frac{h}{a(y)}\frac{d}{dy} \frac{d}{dx}}
~x~
e^{ \mp \frac{h}{a(y)}\frac{d}{dy} \frac{d}{dx}}.
\label{x1}
\eea
Since $e^{\frac{h}{a(y)}\frac{d}{dy} \frac{d}{dx}}$ commutes with $b\frac{d^2}{dx^2}$
we find
\bea
 (a(y)x+h\frac{d}{dy})^2~+~b\frac{d^2}{dx^2}=~
e^{\frac{h}{a(y)}\frac{d}{dy} \frac{d}{dx}}
~[(e^{-\frac{h}{a(y)}\frac{d}{dy} \frac{d}{dx}}
~a(y)~e^{\frac{h}{a(y)}\frac{d}{dy} \frac{d}{dx}}
~x)^2~+~b\frac{d^2}{dx^2}]
e^{-\frac{h}{a(y)}\frac{d}{dy} \frac{d}{dx}}. ~~~~~~~
\label{xx33}
\eea
Hence
\bea
&& F(y)=\int_{-\infty}^{+\infty} dy \int_{-\infty}^{+\infty} dx ~
<y|~<x|~
e^{-[(a(y)x+h\frac{d}{dy})^2~+~b\frac{d^2}{dx^2}~+c(y)]}~
|x>~|y> \nonumber \\
&&  = \int_{-\infty}^{+\infty} dy \int_{-\infty}^{+\infty} dx
<y|~<x|
~~e^{\frac{h}{a(y)}\frac{d}{dy} \frac{d}{dx}}
e^{-[(e^{-\frac{h}{a(y)}\frac{d}{dy} \frac{d}{dx}}
a(y)e^{\frac{h}{a(y)}\frac{d}{dy} \frac{d}{dx}}
x)^2+b\frac{d^2}{dx^2}
+e^{-\frac{h}{a(y)}\frac{d}{dy} \frac{d}{dx}}
~c(y)e^{\frac{h}{a(y)}\frac{d}{dy} \frac{d}{dx}}
]} \nonumber \\
&& e^{-\frac{h}{a(y)}\frac{d}{dy} \frac{d}{dx}} ~~|x> |y>.~~~~~
\label{xy3}
\eea
Now inserting complete set of states we find
\bea
&& F(y) = \int dp_y \int dp'_y \int dp_x \int dp'_x
\int_{-\infty}^{+\infty} dy \int_{-\infty}^{+\infty} dx
<y|p_y>~<x|p_x>
~<p_y|<p_x|e^{\frac{h}{a(y)}\frac{d}{dy} \frac{d}{dx}} \nonumber \\
&& e^{-[(e^{-\frac{h}{a(y)}\frac{d}{dy} \frac{d}{dx}}
a(y)e^{\frac{h}{a(y)}\frac{d}{dy} \frac{d}{dx}}
x)^2+b\frac{d^2}{dx^2}
+e^{-\frac{h}{a(y)}\frac{d}{dy} \frac{d}{dx}}
~c(y)e^{\frac{h}{a(y)}\frac{d}{dy} \frac{d}{dx}}
]} e^{-\frac{h}{a(y)}\frac{d}{dy} \frac{d}{dx}}|p'_x>|p'_y><p'_x|x><p'_y|y> \nonumber \\
&&  = \frac{1}{(2\pi)^2}\int dp_y \int dp'_y \int dp_x \int dp'_x
\int_{-\infty}^{+\infty} dy \int_{-\infty}^{+\infty} dx
e^{iyp_y}~e^{ixp_x}
~<p_y|<p_x|e^{\frac{h}{a(y)}\frac{d}{dy} \frac{d}{dx}} \nonumber \\
&& e^{-[(e^{-\frac{h}{a(y)}\frac{d}{dy} \frac{d}{dx}}
a(y)e^{\frac{h}{a(y)}\frac{d}{dy} \frac{d}{dx}}
x)^2+b\frac{d^2}{dx^2}
+e^{-\frac{h}{a(y)}\frac{d}{dy} \frac{d}{dx}}
~c(y)e^{\frac{h}{a(y)}\frac{d}{dy} \frac{d}{dx}}
]} e^{-\frac{h}{a(y)}\frac{d}{dy} \frac{d}{dx}}|p'_x>|p'_y>e^{-ixp'_x}e^{-iyp'_y}.~~~~~
\label{xy31}
\eea

Unlike the situation in eq. (\ref{pf})
we can now change the $x$ integration variable to $x^\prime$ via
\bea
x = x^\prime - \frac{h}{a(y)}\frac{d}{dy}.
\label{xshift}
\eea
This is because (unlike the left hand side of eq. (\ref{xy3})),
$a(y)$ and $x$ can not be interchanged in the right hand side
of eq. (\ref{xy3}). Hence we can change the $x$ variable to
$x^\prime$ via eq. (\ref{xshift}) which involves a derivative.
With the above change in integration variable the integration
limits for $x^\prime$ remain $\pm \infty$. Under this
change of integration variable one also has $dx~=~dx^\prime$.
With these changes we find from the equation (\ref{xy3})
\bea
&& F(y) = \frac{1}{(2\pi)^2}\int dp_y \int dp'_y \int dp_x \int dp'_x
\int_{-\infty}^{+\infty} dy \int_{-\infty}^{+\infty} dx
e^{iyp_y}~e^{ip_x(x-\frac{h}{a(y)}\frac{d}{dy})}
~<p_y|<p_x|e^{\frac{h}{a(y)}\frac{d}{dy} \frac{d}{dx}} \nonumber \\
&& e^{-[(e^{-\frac{h}{a(y)}\frac{d}{dy} \frac{d}{dx}}
a(y)e^{\frac{h}{a(y)}\frac{d}{dy} \frac{d}{dx}}
(x-\frac{h}{a(y)}\frac{d}{dy}))^2+b\frac{d^2}{dx^2}
+e^{-\frac{h}{a(y)}\frac{d}{dy} \frac{d}{dx}}
~c(y)e^{\frac{h}{a(y)}\frac{d}{dy} \frac{d}{dx}}
]} e^{-\frac{h}{a(y)}\frac{d}{dy} \frac{d}{dx}}|p'_x>|p'_y> \nonumber \\
 && e^{-ip'_x(x-\frac{h}{a(y)}\frac{d}{dy})}e^{-iyp'_y}.~~~~~
\label{xy32}
\eea
Using eq. (\ref{x1}) for the similarity transformation of $(x- \frac{h}{a(y)}\frac{d}{dy})$
we find
\bea
&& \int_{-\infty}^{+\infty} dy \int_{-\infty}^{+\infty} dx ~
<y|~<x|~
e^{-[(a(y)x+h\frac{d}{dy})^2~+~b\frac{d^2}{dx^2}~+c(y)]}~
|x>~|y>= \frac{1}{(2\pi)^2}\int dp_y \int dp'_y \int dp_x \nonumber \\
&&~\int dp'_x \int_{-\infty}^{+\infty} dy \int_{-\infty}^{+\infty} dx
e^{iyp_y}~e^{ip_x(x-\frac{h}{a(y)}\frac{d}{dy})} <p_y|<p_x| e^{-a^2(y)x^2+b\frac{d^2}{dx^2}
+c(y)}|p'_x>|p'_y> \nonumber \\
&& e^{-ip'_x(x-\frac{h}{a(y)}\frac{d}{dy})}e^{-iyp'_y}.~~~~~
\label{shift}
\eea
This concludes our derivation of eq. (\ref{ef3x}), and consequently the special case (\ref{special}).

\subsection{Shift Theorem Involving Matrices and Non-Commuting
Operators in the Exponential}

We next consider the  similarity transformation on the matrices  $x \delta^{ij}$ as follows
\bea
\delta^{ij}x \pm [\frac{h}{A(y)}
[\frac{d}{dy}]]^{ij}=[
e^{\pm \frac{d}{dx}\frac{h}{A(y)} [ \frac{d}{dy}] }
x
e^{\mp \frac{d}{dx}\frac{h}{A(y)} [ \frac{d}{dy}] }]^{ij}~~~~~~~
\label{xn12}
\eea
where $A^{ij}(y)$ is $y$ dependent matrix.

Since $[e^{\frac{d}{dx}\frac{h}{A(y)} [\frac{d}{dy}] }]^{ij}$
commutes with $B\delta^{ij}\frac{d^2}{dx^2}$ we find
\bea
&& [(A(y)x+h[\frac{d}{dy}])^2]^{ij}+\delta^{ij}B\frac{d^2}{dx^2}=
[e^{\frac{d}{dx}\frac{h}{A(y)} [ \frac{d}{dy}] }]^{im} [
(e^{-\frac{d}{dx}\frac{h}{A(y)} [ \frac{d}{dy}] } A(y)
e^{\frac{d}{dx}\frac{h}{A(y)} [ \frac{d}{dy}] } x)^2+B\frac{d^2}{dx^2}]^{ml} \nonumber \\
&& [e^{-\frac{d}{dx}\frac{h}{A(y)} [ \frac{d}{dy}] }]^{lj} .~~~~~~~~
\label{xn5}
\eea
Repeating the same logic as used previously, we obtain
\bea
&& I_{ij}(y)~=
~[\int_{-\infty}^{+\infty} dy \int_{-\infty}^{+\infty} dx <y|~<x|e^{-[(A(y)x+h\frac{d}{dy})^2~+B \frac{d^2}{dx^2}~+
C(y)]}~|x> ~|y>]_{ij} \nonumber \\
&&~=  \frac{1}{(2\pi)^2}[\int_{-\infty}^{+\infty} dy~\int_{-\infty}^{+\infty} dx ~\int dp_y \int dp'_y \int dp_x
\int dp'_x~ e^{iyp_y} <p_y|~e^{ixp_x}e^{-ip_x\frac{h}{A(y)}\frac{d}{dy}}  \nonumber \\
&& <p_x|e^{-[A^2(y)x^2~+~B\frac{d^2}{dx^2}~+C(y)]}|p'_x>~e^{-ixp'_x}e^{ip'_x\frac{h}{A(y)}\frac{d}{dy}}|p'_y>~
e^{-iyp'_y}]_{ij}.
\label{shiftmatrix}
\eea

Since this "derivation" is rather formal and relies on similarity transformations that are not very familiar,
we will now give examples demonstrating the usefulness and validity of the special case  eq. (\ref{special}),
assuming that the integrals define a function which is Taylor "expandable" in a series in $h$.

\section{Some special cases}

In this section we would like to consider the special case
\bea
A[h,y] =\int_{-\infty}^{+\infty} dx ~e^{-(a(y)x+h\frac{d}{dy})^2}~f(y)
\label{ee3}
\eea
We would like to show that
\be
A[h,y]=~
\int_{-\infty}^{+\infty} dx ~e^{-a^2(y)x^2}~f(y) = A[h=0,y].
\ee
To do this we will need to assume that $f(y)$ is such that $A[h,y]$ has a unique Taylor expansion in the variable $h$.

To obtain the Taylor series we will use a theorem for two non-commuting operators $A,~ B$
\bea
&& e^{-(A+B)}~=~e^{-A}~[1~+~\sum_{n=1}^\infty ~(-1)^n~\prod_{i=1}^n~[\int_0^{x_{i-1}} dx_i ~
~e^{x_iA}~B~e^{-x_iA} ]~].
\label{geq1}
\eea

Using eq. (\ref{geq1}) in (\ref{ee3}) we find
\bea
&& ~e^{-(a(y)x+h\frac{d}{dy})^2}~f(y)=~
e^{-A}~ [ 1~ - ~ \int_0^1 dx_1 e^{x_1 A} ~B~e^{-x_1 A}~ \nonumber \\
&& ~+\int_0^1 dx_1 ~ e^{x_1 A} ~B~e^{-x_1 A}~
\int_0^{x_1}~dx_2 e^{ x_2 A} ~B~e^{-x_2A}~ \nonumber \\
&& -\int_0^1 dx_1 ~
e^{x_1 A} ~B~e^{-x_1 A}~
\int_0^{x_1}~dx_2e^{ x_2 A} ~B~e^{-x_2A}~
\int_0^{x_2}~dx_3 e^{ x_3 A} ~B~e^{-x_3A}~ +.....~~~~~~~~~]f(y) \nonumber \\
\label{int1}
\eea
where
\bea
&& (a(y)x+h\frac{d}{dy})^2~=~A+B,~~\nonumber \\
&&~A~=a^2(y)x^2~~~~~~~~~~ \nonumber \\
&&~ B=~2x~a(y)~h \frac{d}{dy} ~+~x~h\frac{da(y)}{dy} ~+~ h^2\frac{d^2}{dy^2}.
\label{cd}
\eea
Integrating $x$ from $-\infty$ to $+\infty$ in eq. (\ref{int1}) we write
\bea
A[h,y] =~
\int_{-\infty}^{+\infty} dx ~e^{-A}~[f(y)~-~I_1[h,y]~+~I_2[h,y]~-~I_3[h,y]~
+~I_4[h,y]~+~......] ~~~~
\label{int2}
\eea
where
\bea
I_n[h,y]~=~
\frac{\int_{-\infty}^{+\infty} dx ~e^{-A}~
[\prod_{i=1}^n~[\int_0^{x_{i-1}} dx_i ~ ~e^{x_iA}~B~e^{-x_iA}] ]f(y)}{
\int_{-\infty}^{+\infty} dx ~e^{-A}}
\label{in}
\eea
with $x_0$ =1 and $n$=1,2,3...etc.   The $I_n$ consist of a finite number of terms  in $h$ from $h^n$ up to $h^{2n}$
Using the expressions for $A$ and $B$ from
eq. (\ref{cd}) and performing the $x$ and $x_i$'s
integrations explicitly in eq (\ref{in}) we can obtain
explicit expressions for all $I_n$. Then assuming that after we perform the integrations we can write
\be
A[h,y]= \sum_{n=0}^\infty  A_n[y] h^n
\ee
we will find that except for $n=0$ all the coefficients in the Taylor series in $h$ are zero.

\subsection{Examples using simple function for $a(y)$ and $f(y)$}

First, we will consider two examples for $a(y)$ and $f(y)$
to demonstrate how eq. (\ref{ee3}) works before giving the result
for general $a(y)$ and $f(y)$.  If we look at each power of $h$ in the
expression of $I_n$, each odd power of $h$  formally
vanishes because it contains an odd integration over $x$ (see eq. (\ref{cd})).

\noindent{\bf Example I}: ~~~$a(y)=y$,~~~~~~~~$f(y)=y$.

Using $a(y)=y$ and $f(y)=y$ in eq. (\ref{in}) we find
\bea
-I_1=h^2\frac{1}{2y},~~~~~~~~~~I_2=-h^2 \frac{1}{2y}-h^4 \frac{19}{24 y^3}
\eea
which gives $-I_1+I_2= -h^4 \frac{19}{24 y^3}$,
independent of terms containing two powers of $h$.
Similarly we find
\bea
-I_3=h^4\frac{19}{12y^3}+h^6 \frac{35}{48y^5},~~~~~~~~~~~~~I_4=-h^4\frac{19}{24y^3}+h^6\frac{123}{48 y^5} +h^8\frac{4199}{640y^7}
\eea
which gives
$-I_1+I_2-I_3+I_4=
h^6\frac{79}{24 y^5} +h^8\frac{4199}{640y^7}$,
independent of terms containing four powers of $h$.
This process can be repeated and we find $-I_1+I_2-.....I_n$ is independent
of terms containing upto $n$ powers of $h$.
Thus if we assume that our series in $B$ gives us the unique Taylor Series in $h$ , then we find that the answer is independent of $h$ which is what we wished to show.

\noindent{\bf Example II}:~~~~~~$a(y)=\frac{1}{y^2}$,~~~~~~~~~~~$f(y)=e^{-y}$.

Using $a(y)=\frac{1}{y^2}$ and $f(y)=e^{-y}$ in eq. (\ref{in}) we find
\bea
-I_1=e^{-y}~h^2[-1+\frac{2}{y}+\frac{1}{y^2}],~~~~~
I_2=e^{-y} \big \{~ h^2[1-\frac{2}{y}-\frac{1}{y^2}]+h^4
[\frac{1}{2}-\frac{2}{y}
-\frac{5}{3y^2} +\frac{5}{6y^4}] \big \},
\eea
which gives $-I_1+I_2= e^{-y}h^4[\frac{1}{2}-\frac{2}{y}
-\frac{5}{3y^2} +\frac{5}{6y^4}]$,
independent of terms containing two powers of $h$ (or $\frac{d}{dy}$).
Similarly we find
\bea
&& -I_3=e^{-y}\big \{~ h^4[-1+\frac{4}{y}
+\frac{20}{3y^2} -\frac{20}{3y^4} ] +h^6[-\frac{1}{6}+\frac{1}{y}
 +\frac{7}{6y^2} -\frac{7}{2y^4} -\frac{7}{y^5} -\frac{29}{6y^6}] \big \}, \nonumber \\
&& I_4=e^{-y} \big \{~h^4[\frac{1}{2}-\frac{2}{y}
-\frac{5}{y^2} +\frac{35}{6y^4} ]
+h^6[\frac{1}{2}-\frac{3}{y}
 -\frac{49}{6y^2} +\frac{91}{2y^4} +\frac{91}{y^5} +\frac{101}{2y^6} ] \nonumber \\
&&+h^8[\frac{1}{24}-\frac{1}{3y}
 -\frac{1}{2y^2} +\frac{15}{4y^4} +\frac{15}{y^5}
+\frac{1313}{42y^6} +\frac{736}{21y^7} +\frac{2855}{168y^8} ]
\big \} \nonumber \\
\eea
which gives
\bea
&&-I_1+I_2-I_3+I_4=
e^{-y} \big \{ h^6[\frac{1}{3}-\frac{2}{y}
 -\frac{7}{y^2} +\frac{42}{y^4} +\frac{84}{y^5} +\frac{137}{3y^6}] \nonumber \\
&&+h^8[\frac{1}{24}-\frac{1}{3y}
 -\frac{1}{2y^2} +\frac{15}{4y^4} +\frac{15}{y^5}
+\frac{1313}{42y^6} +\frac{736}{21y^7} +\frac{2855}{168y^8} ]
\big \} \nonumber \\
\eea
independent of terms containing four powers of $h$ .
So continuing this reasoning to larger $n$ we find again naively the result is independent of $h$.  Of course for the above choices of $a(y)$,  if one wants to also integrate
over $y$ one must exclude the origin in further integration over $y$ for this result to make sense.

\subsection{General $a(y)$ and $f(y)$ }

For general $a(y)$ and $f(y)$ we find from eq. (\ref{in})
\bea
-I_1=h^2\left(-\frac{f[y] a'[y]^2}{2 a[y]^2}+\frac{a'[y] f'[y]}{a[y]}+\frac{f[y] a''[y]}{2 a[y]}-f''[y]\right)
\label{ir1}
\eea
and
\bea
&&I_2=h^2\left(\frac{f[y] a'[y]^2}{2 a[y]^2}-\frac{a'[y] f'[y]}{a[y]}-\frac{f[y] a''[y]}{2 a[y]}+f''[y]\right) + \nonumber \\
&&
h^4\left(\frac{5 f[y] a'[y]^4}{24 a[y]^4}-\frac{a'[y]^3 f'[y]}{a[y]^3}-\frac{f[y] a'[y]^2 a''[y]}{a[y]^3}+\right.
\frac{2 a'[y] f'[y] a''[y]}{a[y]^2}+\frac{3 f[y] a''[y]^2}{8 a[y]^2}+\frac{4 a'[y]^2 f''[y]}{3 a[y]^2}- \nonumber \\
&& \frac{7 a''[y] f''[y]}{6 a[y]}+\frac{7 f[y] a'[y] a^{(3)}[y]}{12 a[y]^2}-\frac{2 f'[y] a^{(3)}[y]}{3 a[y]}-
\left.\frac{a'[y] f^{(3)}[y]}{a[y]}-\frac{f[y] a^{(4)}[y]}{6 a[y]}+\frac{1}{2} f^{(4)}[y]\right)
\label{ir2}
\eea
By adding eqs. (\ref{ir1}) and (\ref{ir2}) we find
\bea
&& -I_1+I_2=h^4
 \left(\frac{5 f[y] a'[y]^4}{24 a[y]^4}-\frac{a'[y]^3 f'[y]}{a[y]^3}-\frac{f[y] a'[y]^2 a''[y]}{a[y]^3}+
\frac{2 a'[y] f'[y] a''[y]}{a[y]^2}+\frac{3 f[y] a''[y]^2}{8 a[y]^2}+\right. \nonumber \\
&& \left. \frac{4 a'[y]^2 f''[y]}{3 a[y]^2}-
 \frac{7 a''[y] f''[y]}{6 a[y]}+\frac{7 f[y] a'[y] a^{(3)}[y]}{12 a[y]^2}-\frac{2 f'[y] a^{(3)}[y]}{3 a[y]}-
\frac{a'[y] f^{(3)}[y]}{a[y]} \right. \nonumber \\
 && \left.
-\frac{f[y] a^{(4)}[y]}{6 a[y]}+\frac{1}{2} f^{(4)}[y] \right) \nonumber \\
\label{ir12}
\eea
which is independent of terms containing two powers of $h$.

Similarly evaluating $I_3$ and $I_4$ we find that  $-I_1+I_2-I_3+I_4$
does not contain terms up to four powers of $h$.

This process can be repeated up to arbitrary powers of $h$  so we find that the coefficients of $h^n$
for $n$=1,2,3.... vanish

Thus if there is a unique Taylor series for  $A[h,y]$, then we obtain
\be
A[h,y] = A[ h=0,y]
\ee
which is what we wanted to show.

\section{Conclusions}

To conclude, we have shown that, remarkably, inside of integrals over the entire real line one can shift the non-commuting derivative operator (not depending on the integration
variable)  which occurs in
exponentials just as if it were a constant.  In particular we have shown that
\bea
&& \int_{-\infty}^{+\infty} dy \int_{-\infty}^{+\infty} dx ~
<y|~<x|~
e^{-[(a(y)x+h\frac{d}{dy})^2~+b\frac{d^2}{dx^2}~+c(y)]}~
|x>~|y>~= \nonumber \\
 && \int_{-\infty}^{+\infty} dy \int_{-\infty}^{+\infty} dx ~<y|~<x-\frac{h}{a(y)}\frac{d}{dy}|~
e^{-[a^2(y)x^2~+~b\frac{d^2}{dx^2}~+c(y)]}~
|x-\frac{h}{a(y)}\frac{d}{dy}>~|y> \nonumber \\
\label{ef3xccxf}
\eea
as well as the extension to Matrix functions where $x$ integration from $-\infty$ to $+\infty$ must be performed
for the above equation to be true. This equation leads to the special case
\bea
I(y)= \int_{-\infty}^{+\infty} dx ~e^{-(a(y)x+h\frac{d}{dy})^2~}~f(y)=\int_{-\infty}^{+\infty} dx ~e^{-a^2(y)x^2}~f(y)
\label{specialff}
\eea
where $h$ and $b$ are constants and $f,~a,~b~c$ are functions of single variable chosen that the integration over $x$ is well defined.
This shift theorem should prove useful in the evaluation of Path Integrals that occur when utilizing the background field method.

\acknowledgments

We thank Warren Siegel, George Sterman and Peter van Nieuwenhuizen for discussions.
This work was supported in part by the National Science Foundation, grants PHY-0354776
and PHY-0345822. One of us (F.C.) would like to thank both the Santa Fe Institute and
the Physics Department at Harvard University for their hospitality while this work was done.

\end{document}